\documentclass[aps,prl,reprint,superscriptaddress,amsmath,amssymb,longbibliography,floatfix]{revtex4-2}
\usepackage{graphicx}
\usepackage{xcolor}
\usepackage[normalem]{ulem}
\usepackage[colorlinks=true,linkcolor=blue!55!black,citecolor=blue!55!black,urlcolor=blue!55!black]{hyperref}
\makeatletter\@booleantrue\raggedcolumn@sw\makeatother %
\begin{document}
\raggedbottom
\brokenpenalty=10000
\clubpenalty=10000
\title{Explosive instability, quiet plasma: phase-space cascades, coherent vortices, and transport suppression in a dipole pair plasma}
\author{D. Kennedy}
\email{daniel.kennedy@ukaea.uk}
\affiliation{UKAEA (United Kingdom Atomic Energy Authority), Culham Campus, Abingdon, Oxfordshire OX14 3DB, United Kingdom}
\author{G. G. Plunk}
\affiliation{Max Planck Institute for Plasma Physics, 17491 Greifswald, Germany}
\begin{abstract}
Turbulence in a magnetised dipole pair plasma can transfer electrostatic energy into coherent flows while carrying entropy to small scales. This occurs via a dual cascade in a four-dimensional phase space, whereby advection by the fluctuating electric field nonlinearly phase-mixes the transit-averaged particle motion. Below a critical pressure gradient, zonal bands suppress the kinetic mode, leaving a transport floor. When the fluid mode is unstable, an explosive burst collapses into a zonal condensate on which two vortices ``breathe'' by reversible energy exchange, without causing net transport.
\end{abstract}
\maketitle

Magnetic confinement of dilute electron--positron plasmas is being pursued by the APEX levitated-dipole and EPOS stellarator programmes~\cite{Pedersen2012,Stoneking2020}. These plasmas offer a simple setting in which to study turbulence under conditions of charge and mass symmetry. When their equilibrium densities and temperatures are also equal, electrostatic drift modes are stable in a straight magnetic field~\cite{Helander2014}. The curvature needed for dipole confinement nevertheless makes the magnetic drift charge dependent, so that density and temperature gradients drive interchange instabilities~\cite{MPH2018,Kennedy2020,Mishchenko2023}. Above a threshold in the gradients (strong drive) this is the fluid interchange; below it (weak drive), a kinetic interchange, the entropy mode~\cite{Simakov2001,Ricci2006}, destabilised by particles resonant with the magnetic drift~\cite{MPH2018}. Linear thresholds are only a first step towards predicting confinement. What matters is how much transport survives once the fluctuations saturate.

In two-dimensional fluids, an inverse energy cascade builds large vortices while enstrophy is transferred to small scales~\cite{Fjortoft1953}. A kinetic plasma can also develop fine structure in the particle distribution, allowing coherent flows to coexist with a forward entropy cascade~\cite{Plunk2010,PlunkTatsuno2011}. In a dipole, particles of different pitch angle follow different orbits through the varying magnetic field. Nonlinear advection can therefore mix the distribution in pitch angle even when perpendicular mixing by Larmor motion~\cite{Dorland1993,Schekochihin2008} is negligible.

We predict and demonstrate a four-dimensional phase-space cascade in which electrostatic energy collects at large scales while entropy develops fine pitch-angle structure. The resulting zonal states exhibit a finite transport floor at weak drive, but transition to vortex breathing at strong drive: the periodic growth and decay of coherent vortices through energy exchange with the zonal flow, accompanied by negligible mean transport.

\begin{figure}[!t]
\centering
\includegraphics{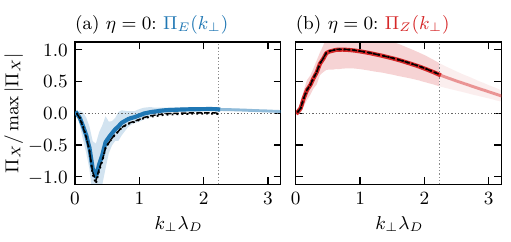}
\caption{(a) The flux $\Pi_E$ of the electrostatic energy $E$ of Eq.~\eqref{eq:E}, and (b) the flux $\Pi_Z$ of the entropy $Z$ of Eq.~\eqref{eq:Z}, through the perpendicular wavenumber $k_\perp$, Eq.~\eqref{eq:flux}, averaged over the nonlinear-onset phase before zonal flows have developed; each is normalised to its own peak. For comparison the black curves use both terms of Eq.~\eqref{eq:E} instead of just its field term; the vertical dotted line marks the largest complete shell of the box.}
\label{fig:cascade}
\end{figure}

\begin{figure}[!t]
\centering
\includegraphics{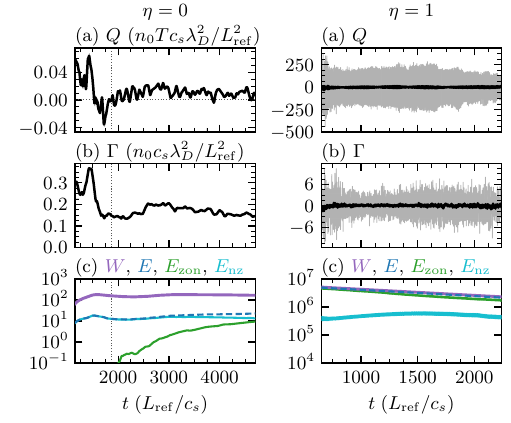}
\caption{(a) Heat flux $Q$ and (b) particle flux $\Gamma$, every output sample (pale) with a short running mean (black). (c) The free energy $W$ of Eq.~\eqref{eq:W}, the electrostatic energy $E$ of Eq.~\eqref{eq:E} with both its terms  and its zonal and non-zonal parts, $E_{\mathrm{zon}}, E_{\mathrm{nz}}$.}
\label{fig:transport_rates}
\end{figure}

\begin{figure}[!t]
\centering
\includegraphics{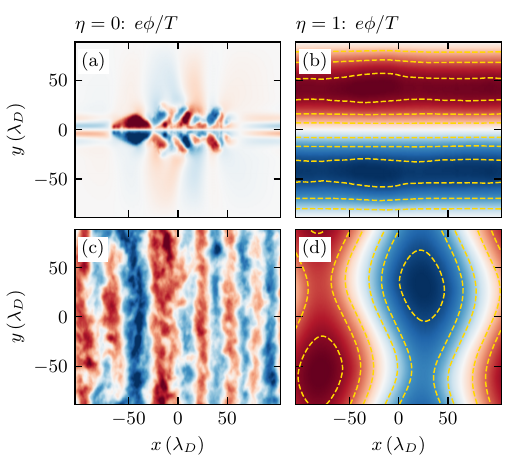}
\caption{(a,c) Electrostatic potential for $\eta=0$ during the nonlinear onset and late in the zonal-band state. (b,d) Same for $\eta=1$ just before the flux peak and late in the drain; the dashed lines are contours of $\phi$, i.e.\ $\mathbf E\times\mathbf B$ streamlines.}
\label{fig:snapshots}
\end{figure}

\textit{Kinetic equations and invariants.---}
We consider the ordering $\rho\ll\lambda_D\ll L$, where $\rho=\sqrt{mT}/(eB)$ is the Larmor radius, $\lambda_D=\sqrt{\epsilon_0T/(2n_0e^2)}$ the Debye length, and $L$ an equilibrium length. The unstable scales, $k_\perp\lambda_D\sim1$, are much larger than $\rho$. This reverses the usual ordering of magnetic confinement, in which $\lambda_D \ll \rho$ and the plasma is quasineutral, $k_\perp\lambda_D \ll 1$, at the unstable Larmor radius scales.

We use flux coordinates $(\psi,\alpha,l)$ with $\mathbf B=\nabla\psi\times\nabla\alpha$ and $l$ the arc length along the field line, and write $\delta f_a=h_a-e_a\phi f_0/T$, with $e_p=-e_e=e$ and a common Maxwellian $f_0$. Below the transit and bounce frequencies, the leading $h_a(\psi,\alpha,\varepsilon,\lambda,t)$ is independent of $l$~\cite{Helander2014,MPH2018}, where $\varepsilon=mv^2/2$ and the pitch-angle variable $\lambda=v_\perp^2/(Bv^2)$. The potential sampled by an orbit is
\begin{equation}
 \begin{aligned}
 \bar\phi(\lambda)&=\frac{1}{\hat\tau(\lambda)}
 \int_{l_1}^{l_2}\frac{\phi(l)\,\mathrm{d} l}{\sqrt{1-\lambda B(l)}},\\
 \hat\tau(\lambda)&=\int_{l_1}^{l_2}\frac{\mathrm{d} l}{\sqrt{1-\lambda B(l)}}.
 \end{aligned}
 \label{eq:orbit}
\end{equation}
The limits are bounce points for trapped particles and one field-line period for passing particles. To leading order in $k_\perp\rho$, the kinetic and field equations are~\cite{Abel2013,MPH2018}
\begin{align}
 \partial_t h_a+\{\bar\phi,h_a\}+\bar v_{da}\partial_\alpha h_a
 &=\frac{e_af_0}{T}(\partial_t\bar\phi-v_{\star a}^{T}\partial_\alpha\bar\phi),\label{eq:kinetic}\\
 (1-\lambda_D^2\nabla_\perp^2)\frac{e\phi}{T}
 &=\frac{1}{2n_0}\int\mathrm{d}^3v\,(h_p-h_e).\label{eq:poisson}
\end{align}
Here $\{X,Y\}=\partial_\psi X\,\partial_\alpha Y-\partial_\alpha X\,\partial_\psi Y$ is the $\mathbf E\times\mathbf B$ advection, $\bar v_{da}$ is the orbit-averaged binormal magnetic drift, and $v_{\star a}^{T}=-(T/e_a)\partial_\psi\ln f_0$ is the gradient drive at fixed $\varepsilon$. The dipole has no radial magnetic drift. 

Let $\left\langle \cdot\right\rangle=V^{-1}\int\mathrm{d}^3r\,(\cdot),$ the average over the flux tube of volume $V,$ and define the field energy $E_D=(n_0e^2/2T)\left\langle \lambda_D^2|\nabla_\perp\phi|^2\right\rangle$. The two nonlinear invariants are
\begin{align}
 W&=\sum_a\left\langle \int\mathrm{d}^3v\,\frac{T\,\delta f_a^2}{4f_0}\right\rangle+E_D,\label{eq:W}\\
 E&=E_D+\frac{n_0e^2}{4TV}\int\mathrm{d}\psi\,\mathrm{d}\alpha\,\mathrm{d}\lambda\;
 \hat\tau(\overline{\phi^2}-\bar\phi^2).\label{eq:E}
\end{align}
Here $W$ is the free energy, the perturbed entropy of the particles weighted by $T$ plus the field energy~\cite{KrommesHu1994}, and $E$ is the electrostatic energy seen by orbit-averaged particles. The second term in Eq.~\eqref{eq:E}, the orbit-variance term, is non-negative~\cite{Helander2013} and vanishes for a flute-like potential, constant along the field line. This $E$ is the electrostatic energy of the linear theory of trapped-particle modes~\cite{Proll2012,Helander2013}; to our knowledge its role as a nonlinear invariant of the orbit-averaged problem has not been examined. The entropy of the orbit-averaged perturbation is
\begin{equation}
 Z=W-E=\sum_a\left\langle \int\mathrm{d}^3v\,\frac{T\,\overline{\delta f}_a^{\,2}}{4f_0}\right\rangle\geq0,
 \label{eq:Z}
\end{equation}
where $\overline{\delta f}_a=h_a-e_a\bar\phi f_0/T$. We call $Z$ entropy throughout. The bracket therefore conserves $Z,$ but the driven dynamics need not. 

The pair $(E,Z)$ constrains transfer between perpendicular Fourier modes, as energy and enstrophy do in two-dimensional fluids~\cite{Fjortoft1953}. At $k_\perp\lambda_D\ll1$, the least entropy a mode carries per unit of $E$ obeys $\min(\hat Z/\hat E)=O(k_\perp^2\lambda_D^2)$, $\hat Z$ and $\hat E$ being spectral densities of $Z$ and $E$, and the state attaining this scaling is flute-like [proof in the End Matter, Eqs.~\eqref{eq:A_measure}--\eqref{eq:A_sandwich}]. Electrostatic energy must therefore collect at large scales whenever entropy moves to small scales~\cite{Plunk2010,PlunkTatsuno2011} [Eq.~\eqref{eq:A_fjortoft}]. That constraint fixes the end state, not the route to it: it holds whether the energy arrives through a constant-flux cascade, as in two-dimensional fluids, or in single nonlocal steps. A flute-like potential has the same orbit average for every pitch angle, so the endpoint of the inverse transfer is also the state in which particles of every pitch angle are advected together and nonlinear pitch-angle mixing (described below) switches off.

\begin{figure}[!t]
\centering
\includegraphics{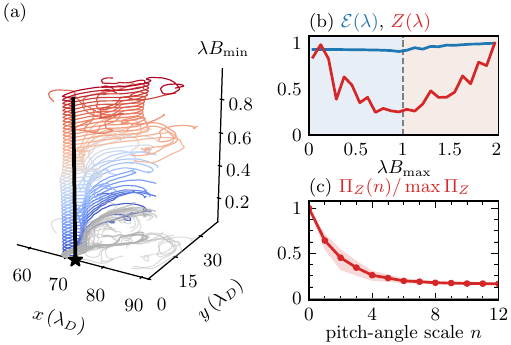}
\caption{(a) Tracers launched at one $(x,y)$ (star), stacked by pitch angle $\lambda B_{\rm min}$ and each advected by its own $\bar\phi(\lambda)$; grey curves are their projections on the floor. (b) Orbit-averaged potential power $\mathcal E(\lambda)$ (blue) and the entropy $Z(\lambda)$ per unit density of states, each normalised to its maximum, over the onset; the dashed line marks the trapped--passing boundary $\lambda=1/B_{\max}$. (c) $\Pi_Z(n)$ over its maximum.}
\label{fig:mixing}
\end{figure}

\textit{Dipole cascade and phase mixing.---}
We use local, electrostatic \texttt{GENE}\ simulations~\cite{Jenko2000} in a circular-coil dipole~\cite{Kennedy2020}, on the flux surface whose outboard equatorial radius is $1.10$ coil radii, so that $|\mathbf B|$ varies by a factor of two along the line. Both drives have $\Omega_\star=\omega_\star/\hat\omega_d=-\mathrm{d}\ln n_0/\mathrm{d}\ln U=1$, with $\hat\omega_d$ the magnetic-drift frequency of a thermal particle and $U=\oint\mathrm{d}l/B$ the specific volume of the flux tube (with $L_{\rm ref}=L_{\rm tube}/2\pi$), and the ratio $\eta = L_n/L_T,$ with $L_n^{-1}=-\partial_x\ln n_0$ and $L_T^{-1}=-\partial_x\ln T$, fixed as $0$ or $1$. The fluid-interchange threshold is $\Omega_\star(1+\eta)=7/4$ in the $Z$-pinch limit~\cite{MPH2018}: at $\eta=0$ the density gradient alone drives the kinetic interchange across a broad band of $k_y$ ($x$ and $y$ are the radial and binormal coordinates of the box, along $\nabla\psi$ and $\nabla\alpha$); at $\eta=1$ the drive is above the threshold and the fluid interchange grows at a scale-independent rate, $\propto k_y / |k_{\perp}|,$ fastest for streamers $(k_x = 0)$. Times are in $L_{\rm ref}/c_s$ with $c_s=\sqrt{T/m}$; fluxes are quoted in the Debye mixing-length units $n_0c_s(\lambda_D/L_{\rm ref})^2$ and $n_0Tc_s(\lambda_D/L_{\rm ref})^2$, energies in $n_0T(\lambda_D/L_{\rm ref})^2$. Both runs use $\lambda_D/\rho=50\sqrt2$ and a box $2L_y$ wide at the equator; fourth-order hyperdiffusion acts in $v_\parallel$ and along the field line and there is no collision operator. Resolution and validation will be detailed in a future work \cite{KennedyPlunkManuscript}.

The nonlinear transfer $T_X^{\rm NL}(\mathbf k)=\partial_t\hat X(\mathbf k)|_{\rm NL}$, from the free-energy diagnostics of \texttt{GENE}~\cite{BanonNavarro2011,BanonNavarro2016}, defines a cumulative flux
\begin{equation}
 \Pi_X(k)=-\sum_{|\mathbf k'|\leq k}T_X^{\rm NL}(\mathbf k'),\qquad X\in\{E,Z\}.
 \label{eq:flux}
\end{equation}
A negative $\Pi_X$ is transfer to larger scales, a positive one to smaller. At $\eta=0$ the two fluxes take opposite signs as soon as the run becomes nonlinear (Fig.~\ref{fig:cascade}): $\Pi_E<0$ across the long-wavelength band and $\Pi_Z>0$ at every $k_\perp.$ This happens before any zonal flow exists: over the averaging window the zonal share of the potential power is negligible, and the store is still filling [Fig.~\ref{fig:transport_rates}(c)]. The condensate then grows out of the turbulence as narrow $k_y=0$ bands, many Debye lengths wide but far below the box scale, with vortices persisting between them [Fig.~\ref{fig:snapshots}(c)]; late in the run they hold most of the equatorial potential power and a large part of $E$, $E_{\mathrm{zon}}$ in Fig.~\ref{fig:transport_rates}(c), still rising, and the transport has not switched off (Fig.~\ref{fig:transport_rates}). At $\eta=1$ there is no turbulent cascade since there is no inertial range to cascade through: the run passes from linear growth to a box-scale zonal flow in one burst, without developed turbulence (Fig.~\ref{fig:snapshots}).

The simulations retain the parallel motion, so their nonlinearity conserves $W$ and $E_D$ but not the orbit-variance term of Eq.~\eqref{eq:E} instantaneously.   However, the full conservation is indeed observed, over timescales longer than the fast parallel dynamics. 

Equation~\eqref{eq:orbit} explains the nonlinear phase mixing. A deeply trapped particle samples the potential near the magnetic-field minimum; a passing particle samples the whole field line. Tracers launched from one point in the $\eta=0$ turbulence, each advected by its own $\bar\phi(\lambda)$, spread to a separation of many Debye lengths [Fig.~\ref{fig:mixing}(a) and Movie 1 \cite{SM}; compare Fig.~1 of Ref.~\cite{Plunk2012}]. Here, the orbit average plays a role analogous to the velocity-dependent gyroaveraging in finite-Larmor-radius turbulence~\cite{Dorland1993,Schekochihin2008}.

Pitch-angle structure produces parallel structure through Eq.~\eqref{eq:poisson}, coupling \(l\) and \(\lambda\) within the four-dimensional phase space \((\psi,\alpha,l,\lambda)\). At \(\eta=0\), the potential power \(\mathcal E(\lambda)=\sum_{\mathbf k}|\bar\phi_{\mathbf k}(\lambda)|^2\) is dominated by nearly flute-like large scales, whereas \(Z(\lambda)\) has order-one variation [Fig.~\ref{fig:mixing}(b)]. Smaller scales retain order-one pitch dependence and separate the tracers. To measure transfer in pitch angle, we expand the orbit-averaged distribution in polynomials in $\lambda$ of degree $n$, orthogonal under the orbit measure $\hat\tau\,\mathrm{d}\lambda$, so that $n$ plays the role of a pitch-angle wavenumber. The flux $\Pi_Z(n)$, the nonlinear transfer of $Z$ out of degrees up to $n$, is positive [Fig.~\ref{fig:mixing}(c)] and entropy moves to finer pitch-angle structure.

\textit{Coherent vortices and breathing.---}
At $\eta=0$ streamers break into small vortices that fill the box (Fig.~\ref{fig:snapshots}). At $\eta=1$ streamers ($k_x = 0$) of every wavelength grow at the same linear rate, while the box-scale streamer dominates in saturation amplitude. A secondary instability collapses the streamer in one burst into box-scale zonal bands modulated by two vortices, one on each jet: above the fluid threshold, here $\Omega_{\star}(1+\eta) = 2 > 7/4,$ the instability does not rely on the drift resonance that phase-mixes the kinetic interchange, so free energy accumulates undissipated and must drain afterwards. The collapse is the two-dimensional secondary instability of the streamer, whose growth rate through the rise of the zonal energy follows the streamer amplitude as $k_y^2|\phi_{k_y}|$ \cite{Rogers2000}. The burst leaves the zonal flow with most of the electrostatic energy $E$. Nothing in the collisionless model damps the flute-like part of this condensate: there is no geodesic curvature, gyroaveraging is negligible at $k_\perp\rho\ll1$, and the velocity-space hyperdiffusion conserves particles, so it cannot change $\phi$. Collisions would damp the flow at a small fraction of the collision frequency, so the condensate is long-lived rather than permanent. In electron-ion plasmas, zonal flows can be driven unstable (a tertiary instability) by a resonance involving nonlinear phase mixing ~\cite{Rogers2000,Ivanov2020,Hallenbert_Plunk_2021}; here, the orbit average could provide such a mechanism through the trapped particles, but none is observed: the condensate persists to the end of the run.

\begin{figure}[!t]
\centering
\includegraphics{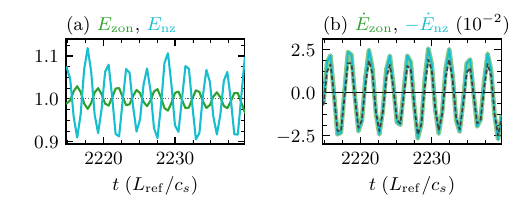}
\caption{(a) The zonal and non-zonal parts of $E$, Eq.~\eqref{eq:E} with both its terms, each over its window mean. (b) The two sides of Eq.~\eqref{eq:exchange} for the field term $E_D$, the first term of Eq.~\eqref{eq:E}.}
\label{fig:breathing}
\end{figure}

In the condensate each vortex gains and loses electrostatic energy through exchange with the zonal flow, and the nonlinear conservation law gives
\begin{equation}
 (\partial_t E_{\rm zon})_{\rm NL}
 =-(\partial_t E_{\rm nz})_{\rm NL},
 \label{eq:exchange}
\end{equation}
where $E_{\rm zon}$ is the $k_y=0$ part of $E$ and $E_{\rm nz}$ the rest. Each vortex, grows and shrinks periodically: it is advected there at a fraction of $k_yU_{\max}$, $U_{\max}$ being the peak zonal flow speed, tilted and untilted by the shear on either side, and the tilting is the exchange of Eq.~\eqref{eq:exchange}. The two vortices move in opposite directions, so in the frame of either the other passes at twice the advection frequency; the tilt of the flow between them, and with it the exchange, follows their relative phase, and the breathing runs at that doubled frequency (Fig.~\ref{fig:breathing} and Movie 2~\cite{SM}).

The large oscillating fluxes of Fig.~\ref{fig:transport_rates}(a,b) are caused by these vortices. Each vortex carries its own density and pressure perturbation outward through one half of its circulation and back through the other, so both fluxes reverse every half turn and average to zero. The free energy injection $P$, Eq.~\eqref{eq:injection}, is zero on average. This is a dynamic analogue of the Dimits state~\cite{Dimits2000}, in which zonal flows quench transport: here the eddies survive, the shear returning the energy it takes.

The zonal shear is a few times the peak linear growth rate at \(\eta=0\), but tens of times that rate at \(\eta=1\) [Figs.~\ref{fig:em_eta0}(c) and \ref{fig:em_eta1}(c)]. The more explosive instability builds the stronger flow and leaves less residual transport.

\begin{figure}[!t]
\centering
\includegraphics{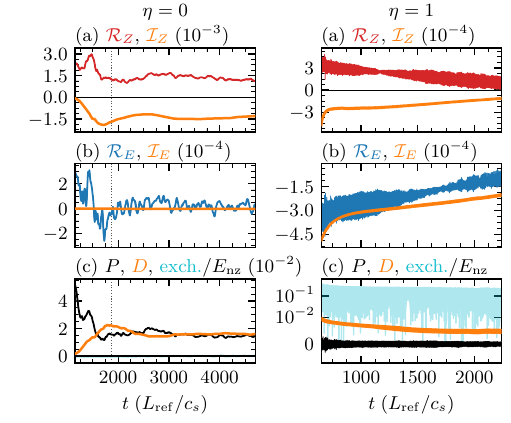}
\caption{(a) $\mathcal{R}_Z$ and $\mathcal{I}_Z$; (b) $\mathcal{R}_E$ and $\mathcal{I}_E$; both normalised by the mean of $W.$ (c) $P, D$ and the breathing exchange, Eq.~\eqref{eq:exchange}, divided by the non-zonal electrostatic energy $E_\mathrm{nz}$.}
\label{fig:energy_rates}
\end{figure}

\textit{Transport and the free-energy balance.---}
Define outward-positive, species-summed radial fluxes
$\Gamma=\sum_a\left\langle v_{E,x}\int\mathrm{d}^3v\,\delta f_a\right\rangle$ and
$Q=\sum_a\left\langle v_{E,x}\int\mathrm{d}^3v\,\varepsilon\delta f_a\right\rangle$.
Here $Q$ is the total heat flux. The total free-energy equation is~\cite{KrommesHu1994,Sugama1996}
\begin{equation}
 \frac{\mathrm{d} W}{\mathrm{d} t}=P-D,\quad
 P\equiv\frac12\left[\frac{T\Gamma}{L_n}
       +\frac{Q-\tfrac32T\Gamma}{L_T}\right].\label{eq:injection}
\end{equation}
Magnetic drift and nonlinear advection conserve $W$; $D$ is its irreversible removal.

Figure~\ref{fig:energy_rates}(a,b) splits the rates of change of $Z$ and $E$ into reversible parts, $\mathcal{R}_{Z}$ and $\mathcal{R}_{E},$ from the gradient drive and the linear terms, and irreversible parts, $\mathcal{I}_{Z}$ and $\mathcal{I}_{E},$ from dissipation, with
$P=\mathcal R_Z+\mathcal R_E$ and $D=-(\mathcal I_Z+\mathcal I_E)$. At $\eta=0,$ over the onset, injection exceeds dissipation and the difference fills the store of energy. When the zonal bands take over the particle flux is smaller but steady, injection and dissipation balance [Fig.~\ref{fig:energy_rates}(c)] and the store stops filling. This is the transport floor of the kinetic interchange, carried by a residual bath of coherent fluctuations. 

At \(\eta=1\), electrostatic energy feeds entropy while the sinks remove free energy [Fig.~\ref{fig:energy_rates}(a,b)]. Mean injection is negligible, so \(\langle dW/dt\rangle_t\simeq-\langle D\rangle_t\), while the reversible zonal–vortex exchange is much faster [Fig.~\ref{fig:energy_rates}(c) and \ref{fig:breathing}]. The state therefore changes little over a breathing cycle but decays over many cycles. Its long-time outcome remains open: dissipation may vanish, leaving a transport-free state, or interchange activity may resume if the zonal shear falls below the stabilising level.

\emph{Confinement implications.---}
For weak drive, the heat flux is small, an order of magnitude below $T\Gamma$ [Fig.~\ref{fig:transport_rates}(a,b)], suggesting the simplifying assumption $\left\langle Q\right\rangle_t=0$.  Steady balance $\left\langle P\right\rangle_t=\left\langle D\right\rangle_t$ then gives $
 T(1-\tfrac32\eta)\left\langle \Gamma\right\rangle_t
 =2L_n\left\langle D\right\rangle_t.$
For $L_n>0$ and positive mean dissipation, particles move inward for $\eta>2/3$ and outward for $\eta<2/3$, the reversal that Helander and Connor~\cite{HelanderConnor2016} found for the quasilinear particle flux near marginal stability. The $\eta=0$ run lies on the predicted side: its particle flux is outward throughout [Figs.~\ref{fig:transport_rates}(b) and~\ref{fig:em_eta0}(a)]. 

At $\eta=1$ the burst moves particles inwards and heat outward [Fig.~\ref{fig:em_eta1}(a)].  An inward pinch has precedent in electron--ion plasmas~\cite{Yankov1994,Isichenko1995,Isichenko1996,Kobayashi2010}, in LDX~\cite{BoxerLDX2010} and RT-1~\cite{Saitoh2011}.

These observations motivate the conjecture that evolving profiles could relax towards $\eta=2/3$. An inward pinch steepens the density profile and lowers $\eta$; outward transport flattens it and raises $\eta$. The proposed endpoint, $T\propto n_0^{2/3}$, has uniform Maxwellian entropy per particle~\cite{LandauLifshitz1980}. The value $\eta=2/3$ forms one boundary of Helander's point-dipole ground-state region~\cite{Helander2017}; the full condition additionally constrains the absolute gradients. It is also the ray on which the fluid- and kinetic-interchange thresholds of the point dipole meet~\cite{HelanderConnor2016, MPH2018}.

The more explosive instability leads to the quieter plasma end state, and the particle fluxes of the two runs point towards $\eta=2/3$ from either side. The strong-drive condensate reaches the box scale, but its size in a device remains undetermined. Global simulations with physical collisions could test its persistence, residual transport, and coupling to the equilibrium. These results concern magnetically confined pair plasmas with $\rho \ll \lambda_D,$ a regime also found in pulsar magnetospheres~\cite{GoldreichJulian1969,PhilippovKramer2022}. The APEX experiment could test these predictions through the evolution of its density and temperature profiles and through coherent potential fluctuations at twice the vortex advection frequency.

\begin{acknowledgments}
\vspace{\baselineskip}\noindent\textit{Acknowledgements.---}We thank J. Ball, R. Nies, A. A. Schekochihin, and W. Clarke for helpful discussions. Part of this work used the CINECA PITAGORA supercomputer under project GEMST26. This work has been funded by the EPSRC Fusion Grant 2022/27 [grant number EP/W006839/1].  To obtain further information on the data and models underlying this paper please contact PublicationsManager@ukaea.uk. For the purpose of open access, the author(s) has applied a Creative Commons Attribution (CC BY) licence (where permitted by UKRI, ‘Open Government Licence’ or ‘Creative Commons Attribution No-derivatives (CC BY-ND) licence’ may be stated instead) to any Author Accepted Manuscript version arising.
\end{acknowledgments}

\bibliography{references,letter-references}

\clearpage

\section*{End Matter}
\setcounter{equation}{0}\renewcommand{\theequation}{A\arabic{equation}}\renewcommand{\theHequation}{A\arabic{equation}}
\section*{Appendix A: Least entropy per unit of electrostatic energy}
We derive bounds on the least entropy a perpendicular Fourier mode must carry per unit of electrostatic energy, restricted to the one-well case of a dipole geometry. States are pairs \(h_p,h_e\), independent of \(l\), with \(\phi\) fixed by Eq. (3).

\textit{Phase-space measure and orbit averages.---}
Let $\langle\!\langle A\rangle\!\rangle=\left\langle\int\mathrm{d}^3v\,A\right\rangle$ denote the phase-space integral per unit volume, so that Eq.~\eqref{eq:Z} reads $Z=\sum_a\langle\!\langle T\overline{\delta f}_a^{\,2}/4f_0\rangle\!\rangle$. In the variables of Eq.~\eqref{eq:orbit}, and summed over the sign of $v_\parallel$, the measure is
\begin{equation}
\mathrm{d}^3r\,\mathrm{d}^3v=2\pi\,\mathrm{d}\psi\,\mathrm{d}\alpha\,v^2\,\mathrm{d}v\,\mathrm{d}\lambda\,\frac{\mathrm{d}l}{\sqrt{1-\lambda B}}.
\label{eq:A_measure}
\end{equation}
The last factor is proportional to the time a particle spends in $\mathrm{d}l$, and with a single well the accessible part of the line at each $\lambda$ is one orbit, so integrating any $A$ over $l$ at fixed $(\psi,\alpha,\varepsilon,\lambda)$ gives $\hat\tau\bar A$, the orbit average of Eq.~\eqref{eq:orbit} applied to $A$. Two consequences of this result allow us to find a useful expression for $E$. Firstly, the phase-space integral of any quantity equals that of its orbit average, $\langle\!\langle A\rangle\!\rangle=\langle\!\langle\bar A\rangle\!\rangle$, so that for $g$ independent of $l$, such as $h_a$ or $f_0$, and for any $A(\mathbf r)$,
\begin{equation}
\langle\!\langle g\phi\rangle\!\rangle=\langle\!\langle g\bar\phi\rangle\!\rangle,\qquad
\langle\!\langle f_0\,\bar A\rangle\!\rangle=n_0\langle A\rangle,
\label{eq:A_identity}
\end{equation}
the second because $\int\mathrm{d}^3v\,f_0=n_0$. Secondly, for $G(\psi,\alpha,\lambda)$ independent of $l$ and $\varepsilon$, $\langle\!\langle f_0G\rangle\!\rangle=(n_0/2V)\int\mathrm{d}\psi\,\mathrm{d}\alpha\,\mathrm{d}\lambda\,\hat\tau G$. With $G=\overline{\phi^2}-\bar\phi^{2}$ and the second identity of Eq.~\eqref{eq:A_identity}, the orbit-variance term of Eq.~\eqref{eq:E} becomes $(e^2/2T)\bigl[n_0\langle\phi^2\rangle-\langle\!\langle f_0\bar\phi^{2}\rangle\!\rangle\bigr]$.

\textit{Electrostatic energy in terms of the orbit-averaged perturbation.---}
To express $E$ through $\overline{\delta f}_a$, take the $\phi$-moment of Eq.~\eqref{eq:poisson}: multiplying by $2n_0e\phi$, averaging over the periodic box and integrating $\phi\nabla_\perp^2\phi$ by parts gives $e\langle\!\langle(h_p-h_e)\phi\rangle\!\rangle=4E_D+(2n_0e^2/T)\langle\phi^2\rangle$. Expanding $\overline{\delta f}_a=h_a-e_a\bar\phi f_0/T$ and using the first identity of Eq.~\eqref{eq:A_identity} to move the orbit average off the potential,
\begin{equation}
\begin{split}
\sum_a e_a\langle\!\langle\overline{\delta f}_a\,\bar\phi\rangle\!\rangle
&=e\langle\!\langle(h_p-h_e)\,\phi\rangle\!\rangle-\tfrac{2e^2}{T}\langle\!\langle f_0\bar\phi^{2}\rangle\!\rangle\\
&=4E_D+\tfrac{2e^2}{T}\bigl[n_0\langle\phi^2\rangle-\langle\!\langle f_0\bar\phi^{2}\rangle\!\rangle\bigr]=4E.
\end{split}
\label{eq:A_E}
\end{equation}

\textit{Lower bound on the entropy.---}
The Cauchy--Schwarz inequality for the measure $\sum_a\langle\!\langle\,\cdot\,\rangle\!\rangle$, applied to $\overline{\delta f}_a/\sqrt{f_0}$ and $e_a\bar\phi\sqrt{f_0}$, bounds the square of Eq.~\eqref{eq:A_E} and gives
\begin{equation}
(4E)^2\le\sum_a\Bigl\langle\!\!\Bigl\langle\frac{\overline{\delta f}_a^{\,2}}{f_0}\Bigr\rangle\!\!\Bigr\rangle\sum_a e_a^2\langle\!\langle f_0\bar\phi^{2}\rangle\!\rangle
=\frac{4Z}{T}\,2e^2\langle\!\langle f_0\bar\phi^{2}\rangle\!\rangle.
\label{eq:A_CS}
\end{equation}
Two dimensionless numbers describe the potential: its mean square perpendicular wavenumber in units of $\lambda_D^{-2}$, and its orbit variance as a fraction of its mean square,
\begin{equation}
\mathcal K=\frac{\langle\lambda_D^2|\nabla_\perp\phi|^2\rangle}{\langle\phi^2\rangle},\qquad
\sigma=1-\frac{\langle\!\langle f_0\bar\phi^{2}\rangle\!\rangle}{n_0\langle\phi^2\rangle}.
\label{eq:A_Ksigma}
\end{equation}
The orbit-variance term of Eq.~\eqref{eq:E} is non-negative, so $\sigma\ge0$, with $\sigma=0$ only for a flute-like potential. With $\mathcal K$ and $\sigma$, Eq.~\eqref{eq:E} reads $E=(n_0e^2/2T)\langle\phi^2\rangle(\mathcal K+\sigma)$, which is positive unless $\phi$ is constant. Hence also $\sigma<1$: at $\sigma=1$ the orbit average $\bar\phi$ vanishes at every pitch angle, so the product in Eq.~\eqref{eq:A_E}, and with it $E$, would vanish. Equation~\eqref{eq:A_CS} then becomes
\begin{equation}
\frac{Z}{E}\ge\frac{\mathcal K+\sigma}{1-\sigma}\ge\mathcal K.
\label{eq:A_bound}
\end{equation}
Equality holds in the first step only if $\overline{\delta f}_a=c\,e_a\bar\phi f_0$ with a single constant $c$ for both species and all of phase space, and in the second only if $\sigma=0$.

\textit{The bound for a single perpendicular mode.---}
A single Fourier mode obeys the same bound. Equation~\eqref{eq:poisson} is linear, so a single pair $\pm\mathbf k$, with $\mathbf k=(k_\psi,k_\alpha)$ conjugate to $(\psi,\alpha)$, is a state on its own. In the local flux tube, whose metric and $B$ depend on $l$ alone, $Z$ and $E$ are sums over $\mathbf k$ of their parts $\hat Z(\mathbf k)$ and $\hat E(\mathbf k)$, so Eq.~\eqref{eq:A_bound} holds for each mode with $Z$ and $E$ replaced by $\hat Z(\mathbf k)$ and $\hat E(\mathbf k)$. For such a state the averages over $(\psi,\alpha)$ in Eq.~\eqref{eq:A_Ksigma} reduce, because $\mathrm{d}^3r=\mathrm{d}\psi\,\mathrm{d}\alpha\,\mathrm{d}l/B$, to line averages, $\langle\cdot\rangle\to\oint(\mathrm{d}l/B)(\cdot)\big/\!\oint\mathrm{d}l/B$, of $|\phi_{\mathbf k}|^2$ and $k_\perp^2\lambda_D^2|\phi_{\mathbf k}|^2$, where $\phi_{\mathbf k}(l)$ is the amplitude and $k_\perp(l)=|k_\psi\nabla\psi+k_\alpha\nabla\alpha|$ varies along the line with the metric. Writing $\kappa(l)= k_\perp^2(l)\lambda_D^2$, Eq.~\eqref{eq:A_Ksigma} becomes
\begin{equation}
\mathcal K_{\mathbf k}=\frac{\oint(\mathrm{d}l/B)\,\kappa|\phi_{\mathbf k}|^2}{\oint(\mathrm{d}l/B)\,|\phi_{\mathbf k}|^2},\qquad
\sigma_{\mathbf k}=1-\frac{\langle\!\langle f_0|\bar\phi_{\mathbf k}|^2\rangle\!\rangle}{n_0\langle|\phi_{\mathbf k}|^2\rangle}.
\label{eq:A_Kk}
\end{equation}
Here $\mathcal K_{\mathbf k}$ is a weighted average of $\kappa$ along the line, and so lies between $\min_l\kappa$ and $\max_l\kappa$.

\textit{Attainment of the bound.---}
Equality throughout Eq.~\eqref{eq:A_bound} requires $\sigma=0$, so that $\bar\phi=\phi$ is constant along the line, and $\overline{\delta f}_a=c\,e_a\phi f_0$, so that $h_a=(c+1/T)\,e_a\phi f_0$; Eq.~\eqref{eq:poisson} at each $l$ then fixes $c+1/T$ to $[1+\kappa(l)]/T$, which a single constant $c$ satisfies only if $k_\perp$ is constant along the line. On such a line the bound is attained, $\min(\hat Z/\hat E)=k_\perp^2\lambda_D^2$ exactly: the analogue of the enstrophy-to-energy ratio $k^2$ of two-dimensional flow, with $\lambda_D$ in the place of the Larmor radius of gyrokinetics~\cite{Plunk2010}. In the dipole $k_\perp$ varies along the line and no state attains equality throughout Eq.~\eqref{eq:A_bound}. Equation~\eqref{eq:A_bound} with $\mathcal K_{\mathbf k}\ge\min_l\kappa$ still bounds the minimum of $\hat Z/\hat E$ from below, and one trial state bounds it from above. The pitch-angle-independent perturbation $h_a=(e_a/e)\,\chi f_0$ in the mode $\mathbf k$, with $\chi$ a constant amplitude, has the potential $e\phi_{\mathbf k}/T=\chi/(1+\kappa)$ by Eq.~\eqref{eq:poisson}, so $\overline{\delta f}_a=(e_a/e)\,\chi f_0\,\bar A$ with $A=\kappa/(1+\kappa)$. Then $\bar A^{2}\le\overline{A^{2}}$ and the second identity of Eq.~\eqref{eq:A_identity} bound the entropy of Eq.~\eqref{eq:Z} from above, while the field term of Eq.~\eqref{eq:E} bounds the electrostatic energy from below,
\begin{equation}
\begin{split}
\hat Z&\le\tfrac12 n_0T\chi^2\bigl\langle\kappa^2(1+\kappa)^{-2}\bigr\rangle,\\
\hat E&\ge\hat E_D=\tfrac12 n_0T\chi^2\bigl\langle\kappa(1+\kappa)^{-2}\bigr\rangle,
\end{split}
\label{eq:A_trialZE}
\end{equation}
and their ratio is an average of $\kappa$ with the positive weight $\kappa/(1+\kappa)^2$, so the trial perturbation gives
\begin{equation}
\frac{\hat Z}{\hat E}\le\frac{\langle\kappa^2(1+\kappa)^{-2}\rangle}{\langle\kappa(1+\kappa)^{-2}\rangle}\le\max_l\kappa.
\label{eq:A_trial}
\end{equation}
Since Eqs.~\eqref{eq:A_bound} and~\eqref{eq:A_Kk} give $\hat Z/\hat E\ge\mathcal K_{\mathbf k}\ge\min_l\kappa$ for every state, for every mode and at any $k_\perp\lambda_D$,
\begin{equation}
\min_l k_\perp^2\lambda_D^2\le\min\frac{\hat Z}{\hat E}\le\max_l k_\perp^2\lambda_D^2,
\label{eq:A_sandwich}
\end{equation}
the minimum being over all states of the mode, $\mathbf k\neq0$, with $\hat E>0$. It is attained by $h_a=(1+F)\,e_af_0\bar\phi/T$, equality in the first step of Eq.~\eqref{eq:A_bound}, with $F$ the least eigenvalue of $(1+\kappa)\phi=(1+F)\,n_0^{-1}\!\int\!\mathrm{d}^3v\,f_0\bar\phi$. Along the simulated flux tube $k_\perp$ varies by a factor of order unity, so the minimum is $O(k_\perp^2\lambda_D^2)$. Equation~\eqref{eq:A_bound} also gives $\hat Z/\hat E\ge\sigma_{\mathbf k}$, so a state with $\hat Z/\hat E=O(k_\perp^2\lambda_D^2)$ has $\sigma_{\mathbf k}=O(k_\perp^2\lambda_D^2)$: its potential departs from a flute-like one by $O(k_\perp\lambda_D)$ in root mean square.

\textit{Consequence for the cascade.---}
The bracket in Eq.~\eqref{eq:kinetic} conserves $E$ and $Z$ separately, and Eq.~\eqref{eq:A_sandwich} says where it can put $E$. Let $F(\mathbf k)$ be the minimum in Eq.~\eqref{eq:A_sandwich}, a property of the mode alone lying between $\min_l\kappa$ and $\max_l\kappa$, so that $\hat Z\ge F\hat E$ for every state. Summing over modes gives $\sum_{\mathbf k}F(\mathbf k)\hat E(\mathbf k)\le Z$: for any threshold $F_{\rm c}$, the electrostatic energy at scales with $F\ge F_{\rm c}$ can never exceed $Z/F_{\rm c}$, while entropy above the bound is not fixed by $E$, and the bound places no constraint on the direction in which it moves. This inequality applied to a three-scale transfer gives the kinetic analogue of Fj\o rtoft's argument~\cite{Fjortoft1953, Plunk2012}. Take three groups of modes with $F=F_0<F_1<F_2$, all of $E$ initially at $F_1$, and let the nonlinearity send electrostatic energies $\hat E_0$ and $\hat E_2$ into the groups $F_0$ and $F_2$. Conserving $E$ and $Z$, with $\hat Z\ge F\hat E$ in each group, gives
\begin{equation}
(F_2-F_1)\,\hat E_2\le(F_1-F_0)\,\hat E_0+Z_{\rm ex},
\label{eq:A_fjortoft}
\end{equation}
where $Z_{\rm ex}=Z-F_1E\ge0$ is the entropy initially above the bound, zero for a state at it. Doubling $\mathbf k$ multiplies $\kappa(l)$ by four at every $l$, so it multiplies $F$ by exactly four where $k_\perp$ is constant along the line, $F=\kappa$; in the dipole it multiplies only the bounds $\min_l\kappa$ and $\max_l\kappa$ of Eq.~\eqref{eq:A_sandwich} by four. With $F_0=F_1/4$ and $F_2=4F_1$, Eq.~\eqref{eq:A_fjortoft} reads $\hat E_2\le\hat E_0/4+Z_{\rm ex}/(3F_1)$: a state at the bound sends at most a quarter as much electrostatic energy an octave towards smaller scales as an octave towards larger. In the dipole Eq.~\eqref{eq:A_fjortoft} holds as written, with the ratio $(F_1-F_0)/(F_2-F_1)$ in place of $1/4$, and the octave estimate is exact only for constant $k_\perp$. Equation~\eqref{eq:A_fjortoft} restricts which transfers are possible rather than predicting which occur: once $Z_{\rm ex}$ is comparable with $(F_2-F_1)E$ it allows either direction; Fig.~\ref{fig:cascade} shows the one taken. Two-dimensional fluid turbulence, with enstrophy exactly $k^2$ times energy, is the case of equality~\cite{Fjortoft1953}.

\section*{Appendix B: Time histories of the two runs}
\begin{figure}[!tb]
\centering
\includegraphics{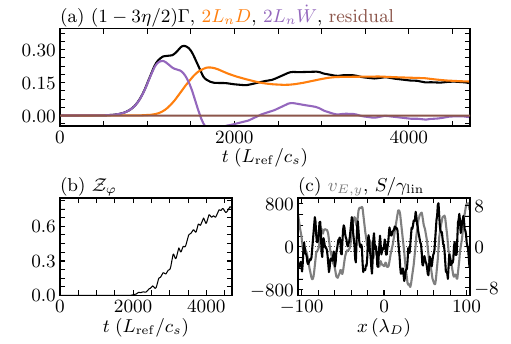}
\caption{The $\eta=0$ run. (a) The terms of Eqs.~\eqref{eq:injection} times $2L_n$, and their residual. (b) The zonal fraction of the equatorial potential power. (c) At the end of the run, the $y$-averaged $\mathbf E\times\mathbf B$ flow $v_{E,y}$ (grey) and the zonal shearing rate $S=\partial_x v_{E,y}$ (black) over the peak linear growth rate $\gamma_{\rm lin}=0.006\,c_s/L_{\rm ref}$ (dotted, $|S|=\gamma_{\rm lin}$).}
\label{fig:em_eta0}
\end{figure}

\begin{figure}[!tb]
\centering
\includegraphics{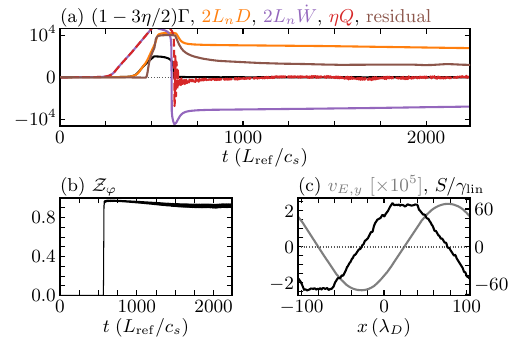}
\caption{As Fig.~\ref{fig:em_eta0}, for the $\eta=1$ run; (a) on a symmetric logarithmic axis, with $\eta Q$ (red, dashed); $\gamma_{\rm lin}=0.0228\,c_s/L_{\rm ref}$ in (c). There are two vortices because the box-scale zonal flow is one jet in each direction, and exchanging the species with $\phi\to-\phi$, $y\to-y$ and a half-box shift, a symmetry of the pair-plasma equations kept by the simulated state, carries each jet and its vortex onto the other.}
\label{fig:em_eta1}
\end{figure}

Figures~\ref{fig:em_eta0} and~\ref{fig:em_eta1} show both runs over their full length.

\end{document}